\documentclass[aps,prl,twocolumn,superscriptaddress,showpacs]{revtex4-1}

\usepackage{graphicx}
\usepackage{hyperref}

\begin{document}


\title{Direct Evidence for Octupole Deformation in $^{146}$Ba and the 
Origin of Large $E1$~Moment Variations in Reflection-Asymmetric Nuclei}

\author{B. Bucher}
\email[]{brian.bucher@inl.gov}
\affiliation{Lawrence Livermore National Laboratory, Livermore,
California 94550, USA}
\affiliation{Idaho National Laboratory, Idaho Falls, Idaho 83415, USA}
\author{S. Zhu}
\email[]{zhu@anl.gov}
\affiliation{Argonne National Laboratory,
Argonne, Illinois 60439, USA}
\author{C. Y. Wu}
\affiliation{Lawrence Livermore National Laboratory, Livermore,
California 94550, USA}
\author{R. V. F. Janssens}
\affiliation{Argonne National Laboratory, Argonne, Illinois 60439,
USA}
\author{R.~N.~Bernard}
\author{L.~M.~Robledo}
\author{T.~R.~Rodr\'{i}guez}
\affiliation{Departamento de F\'{i}sica Te\'{o}rica, Universidad
Aut\'{o}noma de Madrid, E-28049 Madrid, Spain}
\author{D.~Cline}
\author{A. B. Hayes}
\affiliation{University of Rochester, Rochester, New York 14627,
USA}
\author{A.~D.~Ayangeakaa}
\affiliation{Argonne National Laboratory, Argonne, Illinois 60439,
USA}
\author{M.~Q.~Buckner}
\affiliation{Lawrence Livermore National Laboratory, Livermore,
California 94550, USA}
\author{C. M. Campbell}
\affiliation{Lawrence Berkeley National Laboratory, Berkeley,
California 94720, USA}
\author{M.~P.~Carpenter}
\affiliation{Argonne National Laboratory, Argonne, Illinois 60439,
USA}
\author{J.~A. Clark}
\affiliation{Argonne National Laboratory, Argonne,
Illinois 60439, USA}
\author{H.~L.~Crawford}
\affiliation{Lawrence Berkeley National Laboratory, Berkeley,
California 94720, USA}
\author{H. M. David}
\altaffiliation[Present address: ]{GSI Helmholtzzentrum f\"{u}r
Schwerionenforschung, 64291 Darmstadt, Germany} \affiliation{Argonne
National Laboratory, Argonne, Illinois 60439, USA}
\author{C. Dickerson}
\affiliation{Argonne National Laboratory, Argonne, Illinois 60439,
USA}
\author{J. Harker}
\affiliation{Argonne National Laboratory, Argonne,
Illinois 60439, USA} \affiliation{University of Maryland, College
Park, Maryland 20742, USA}
\author{C.~R.~Hoffman}
\author{B.~P.~Kay}
\author{F.~G.~Kondev}
\affiliation{Argonne National Laboratory, Argonne, Illinois 60439,
USA}
\author{T. Lauritsen}
\affiliation{Argonne National Laboratory, Argonne, Illinois 60439,
USA}
\author{A.~O.~Macchiavelli}
\affiliation{Lawrence Berkeley National Laboratory, Berkeley,
California 94720, USA}
\author{R.~C. Pardo}
\affiliation{Argonne National Laboratory, Argonne, Illinois 60439,
USA}
\author{G. Savard}
\affiliation{Argonne National Laboratory, Argonne, Illinois 60439,
USA}
\author{D. Seweryniak}
\affiliation{Argonne National Laboratory, Argonne, Illinois 60439,
USA}
\author{R. Vondrasek}
\affiliation{Argonne National Laboratory, Argonne, Illinois 60439,
USA}


\date{\today}

\begin{abstract}
Despite the more than one order of magnitude difference between the
measured dipole moments in $^{144}$Ba and $^{146}$Ba, the strength
of the octupole correlations in $^{146}$Ba are found to
be as strong as those in $^{144}$Ba with a similarly large value of 
$B(E3;3^- \rightarrow 0^+)$ determined as 48($^{+21}_{-29}$) W.u. 
The new results not only
establish unambiguously the presence of a region of octupole
deformation centered on these neutron-rich Ba isotopes, but also
manifest the dependence of the electric dipole moments on the
occupancy of different neutron orbitals in nuclei with enhanced
octupole strength, as revealed by fully microscopic calculations.
\end{abstract}


\maketitle

Unlike the electrons in atoms, protons and neutrons are closely
bound together in nuclei by the strong nuclear force, occupying
quantum levels that can result in
different nuclear shapes because of sizeable long range
multipole-multipole interactions. The studies of these shapes, and
of the associated nuclear moments, facilitate our understanding of
the origin of simple patterns in such complex quantum many-body
systems. Certain isotopes are thought to develop octupole
deformation due to strong octupole-octupole interactions present
when both types of valence nucleons occupy pairs of single-particle
orbitals near the Fermi surface with orbital ($\ell$) and total
($j$) angular momenta differing by 3$\hbar$ \cite{Butler1996}.
There is now experimental evidence to suggest that nuclei with a
low-lying negative-parity band of states interleaved with the
ground-state positive-parity band and linked by fast $E1$
transitions between the two sequences result from strong octupole
correlations. However, because $E3$ transitions are fundamentally
hindered in the electromagnetic decay of nuclear states when
competing with $E1$ and $E2$ transitions, the presence of strong 
octupole correlations is often inferred from the observation of 
large $E1$ transition probabilities.  The latter are related to 
the intrinsic electric dipole moment and are typically obtained 
from $E1/E2$ intensity ratios, with the $E2$ transition probabilities 
then being estimated from lifetime measurements of low-spin states or 
from systematics.  A direct experimental
determination of the electric octupole moment requires the use of
the Coulomb excitation process for the nuclei of interest.

The neutron-deficient radium isotopes around $^{224}$Ra and the
neutron-rich barium isotopes centered at $^{146}$Ba have been
predicted to belong to the two regions with the strongest octupole
correlations \cite{Butler2016}.
However, large fluctuations, by as much as two orders of magnitude 
for a given spin, in the value of the 
intrinsic electric dipole moment have been well documented in these
two regions \cite{Ahmad1993,Butler1996}, even though other 
spectroscopic features; i.e., negative-parity bands located at 
comparably low excitation energies, strongly suggest the presence 
of similar octupole strengths.
Classically, octupole-deformed nuclei should be
characterized by large electric dipole moments proportional to the
strength of the octupole correlations \cite{Strutinsky1956,Bohr1957}
because of the redistribution of the mass and charge of the protons
and neutrons. Interestingly, in the radium isotopes, a minimum
occurs in the value of the intrinsic electric dipole moment for
$^{224}$Ra, but the magnitude of octupole strength in this nucleus,
as recently quantified through Coulomb excitation with a $^{224}$Ra
radioactive beam, is one of the largest in the region
\cite{Gaffney2013}. Neutron-rich barium nuclei form another
interesting set in terms of studying the relationship between the
intrinsic electric dipole and octupole moments
\cite{Butler1996,Phillips1986,Mach1990}. Specifically, between
$^{144}$Ba$_{88}$ and $^{146}$Ba$_{90}$, the electric dipole moments
are observed to drop suddenly by more than an order of magnitude
\cite{Phillips1986,Mach1990,Urban1997}, but the value quickly
returns to an enhanced level in $^{148}$Ba$_{92}$ \cite{Urban1997}.
Furthermore, it has been pointed out in Ref. \cite{Urban1997} that
the octupole strength in $^{146}$Ba may in fact be quenched as this
could account for the presence of a particle alignment at moderate
angular momentum ($I^\pi$$\sim$12$^+$) in the ground-state band.
Clarification of these issues requires a direct measurement of the $E3$
strength in this nucleus, a challenge until recently because of its
short half-life ($T_{1/2}=2.2$~s~\cite{Peker1997}).

Despite the many experimental challenges associated with 
measurements of the electric octupole moments in these nuclei, there
is also additional motivation in fundamental physics to understand
the relation between the intrinsic electric dipole and octupole
moments: The existence of an atomic electric dipole moment (EDM) has
important implications for $CP$-violation in the early universe that 
could possibly be responsible for the observed asymmetry between
matter and anti-matter \cite{Liu2007}, herewith signifying new
physics beyond the Standard Model \cite{Engel2013}. In diamagnetic
atoms, a measureable EDM could be induced by the so-called Schiff
moment, a quantity that can be enlarged by orders of magnitude by a
sizeable octupole moment and is sensitive to details of the charge
distribution \cite{Auerbach1996,Spevak1997}.  Moreover, the
contribution of the nuclear intrinsic electric dipole moment to the
Schiff moment is not negligible \cite{Spevak1997}. Hence, it is
important to recognize the origin and magnitude of nuclear intrinsic
moments, especially the electric dipole moment as it is closely
associated with the nuclear mass and charge distributions. An
accurate estimate of Schiff moments for different nuclei is required
to evaluate the precision of calculations of this quantity and to
compare the limits on $P$ and $T$ violation reported by various
experiments involving them.

\begin{figure}
   \includegraphics[width=20pc]
{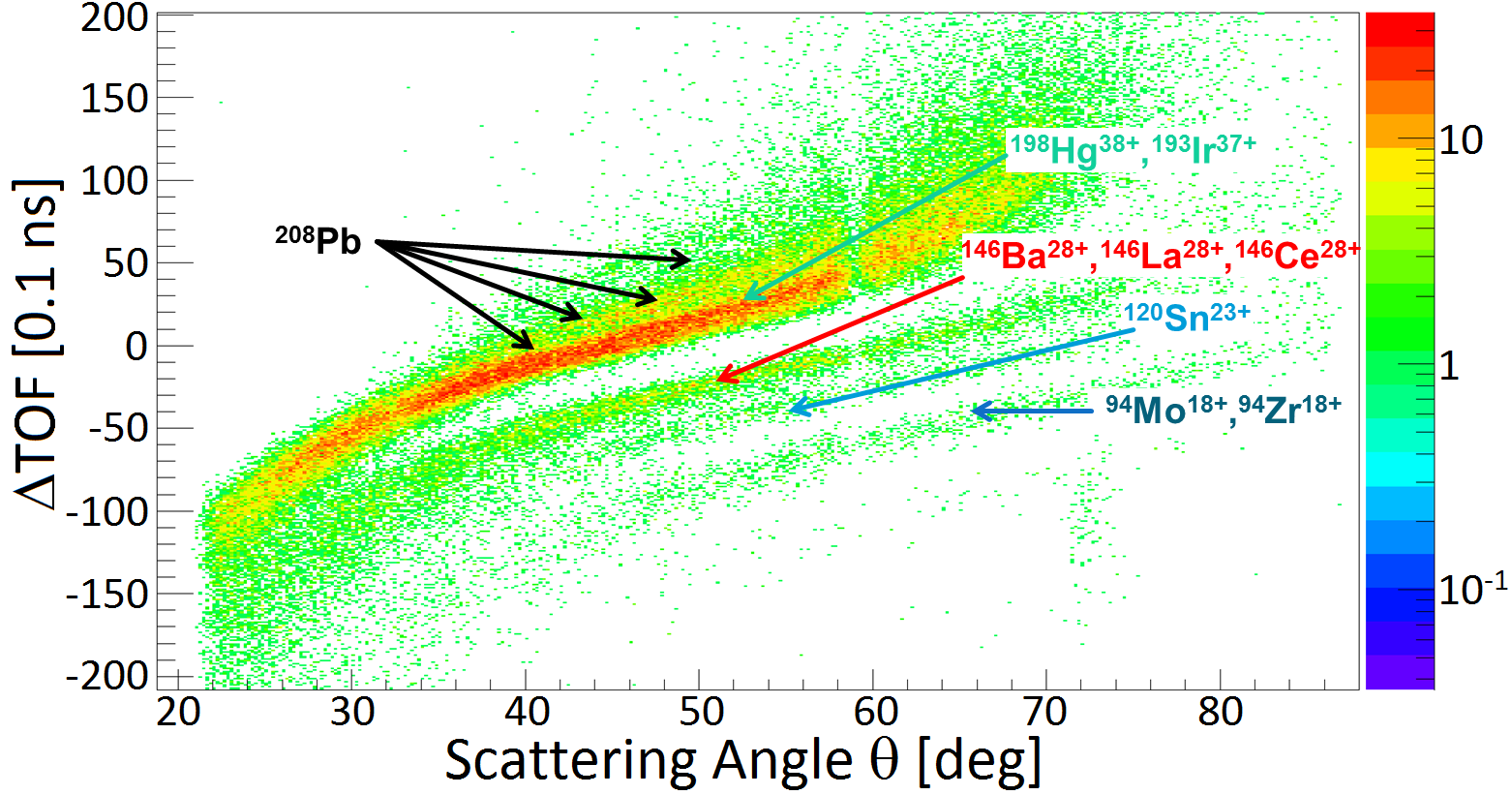}
    \caption{(Color) TOF vs. scattering angle recorded by CHICO2
in coincidence with a $\gamma$~ray observed in GRETINA.  The $A$=146
group is readily distinguished from the stable beam contaminants.}
    \label{fig:TOF}
\end{figure}

\begin{figure}
    \includegraphics[width=20pc]
{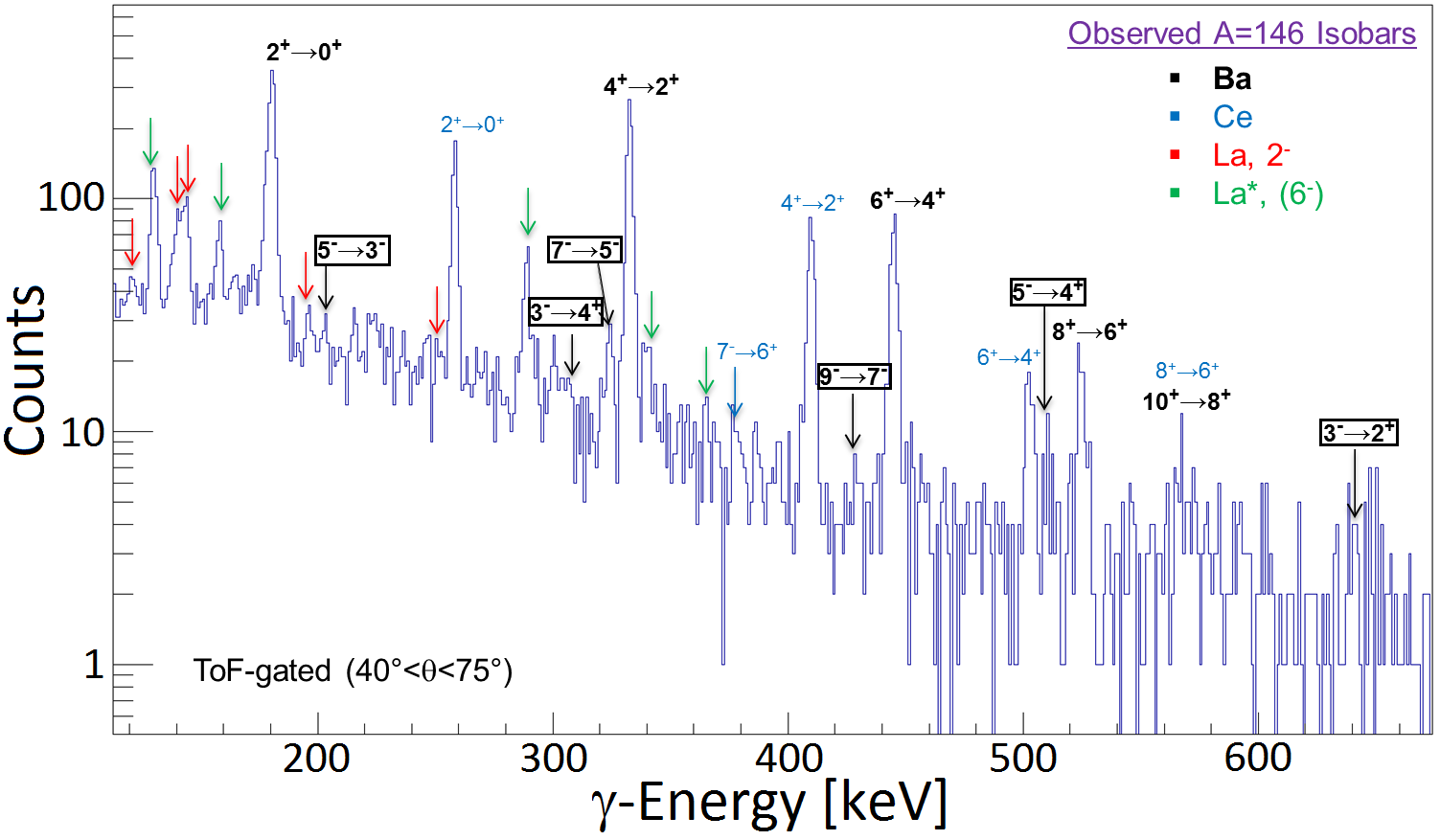}
    \caption{(Color online)
The coincident $\gamma$-ray spectrum obtained by gating on the $A$=146
group in the CHICO2 TOF spectrum (Fig.~\ref{fig:TOF}).  Many
$^{146}$Ba transitions are seen along with lines from the
radioactive isobaric contaminants that were also produced in CARIBU.}
    \label{fig:gam}
\end{figure}

To determine the octupole strength in $^{146}$Ba, a Coulomb
excitation experiment was performed similar to the one carried out
recently for $^{144}$Ba \cite{Bucher2016}. The beam of $^{146}$Ba
ions was produced from $^{252}$Cf fission in the CARIBU facility
\cite{Savard2008,Savard2015} along with the isobaric contaminants
$^{146}$La and $^{146}$Ce, and was charge bred to $q$=28$^+$.  The
$A$=146 beam was accelerated through the ATLAS accelerator to
659~MeV and was focused onto a 1.1~mg/cm$^2$ $^{208}$Pb target
(99.86\% enriched).  The average $^{146}$Ba beam intensity was
3$\times 10^3$ ions per second over 12 days. Additional stable
contaminants (with the same $A/q$) included $^{94}$Mo$^{18+}$,
$^{94}$Zr$^{18+}$, $^{120}$Sn$^{23+}$, $^{193}$Ir$^{37+}$, and
$^{198}$Hg$^{38+}$,  but all were readily separated from the $A$=146
beam, based on time-of-flight (TOF) and scattering angle data
recorded in the CHICO2 heavy-ion
counter~\cite{Wu2016}~(Fig.~\ref{fig:TOF}). This allowed for a clean
$A$=146 $\gamma$-ray spectrum resulting from Coulomb excitation
(Fig.~\ref{fig:gam}).

In the spectrum of Fig.~\ref{fig:gam}, measured with the GRETINA
$\gamma$-ray tracking array~\cite{Paschalis2013}, several
transitions from $^{146}$Ba are apparent, especially those belonging
to positive-parity levels in the ground-state band which are excited
(and decay) by $E2$ transitions. The negative-parity levels are
populated less frequently, but the excitation occurs predominantly
through $E3$ transitions and their decay yields provide a
measurement of the corresponding $E3$ matrix elements.

The $\gamma$-ray yields were extracted for two separate scattering
angle ($\theta$) ranges: 
30$^\circ$--40$^\circ$ and 40$^\circ$--75$^\circ$. At
lower angles, it is difficult to isolate the $A$=146 ions from other
beam contaminants in the TOF spectrum while, at higher $\theta$
values, statistics are insufficient (the cross sections fall off
with the $1/\sin^4 (\frac{\theta}{2})$ Rutherford angular
dependence). The data for each set of angles were analyzed with the
Coulomb excitation least-squares search code, \textsc{gosia}
\cite{Czosnyka1983,*GosiaManualURL}.  Yields were determined for
levels up to 10$\hbar$ in the ground-state band and 9$\hbar$ in the
negative-parity sequence.  The energies of all observed
$\gamma$~rays, along with several branching ratios and level
lifetimes, were known from previous works
\cite{Peker1997,Urban1997,Mach1990,Phillips1986}. The latter
information proved useful for constraining the \textsc{gosia} fit.
The sets of $E1$, $E2$, and $E3$ matrix elements between levels with
no previously known lifetimes were coupled according to the
rigid-rotor prescription \cite{Bohr1975} governed by the individual
intrinsic $E\lambda$ moments (see also
Refs.~\cite{Wollersheim1993,Bucher2016,Phillips1986}). Once the
$\chi^2$ minimum was found, the rigid-rotor constraint was removed
to properly determine the associated uncertainties, including
correlations between matrix elements.  In most cases, the
uncertainty was primarily limited by the lack of statistics in the
measured yields due to the low radioactive beam intensity.

As anticipated, the extracted $E1$ matrix elements did not display
much sensitivity to the data; as mentioned above, the dipole
strength was known to be small from earlier work
\cite{Phillips1986,Mach1990,Urban1997} and, in fact, the only
observed $\gamma$~rays from $E1$ decays in the present measurement
came from the 3$^-$ and 5$^-$ states.
Moreover, the relative sign between the intrinsic $E1$ and $E3$ moments
was found to also be insensitive to the data.  On the other hand,
a number of new $E2$ and $E3$ matrix
elements were well-determined from the data (Table~\ref{tab:me}).

The most significant result obtained here is the ground-state $E3$
matrix element $|\langle 3_1^- \| \hat{M}_{E3} \| 0_1^+ \rangle|$;
it is determined to be 0.65($^{+14}_{-20}$) $e$b$^{3/2}$ and
reflects the amplitude of octupole deformation present in the ground
state \cite{Butler1996}. This value corresponds to a
$B(E3;3^-\rightarrow 0^+)$ reduced transition probability of
48($^{+21}_{-29}$)~W.u. which is essentially the same as the value
of 48($^{+25}_{-34}$)~W.u. reported recently for $^{144}$Ba
\cite{Bucher2016}. This new result supports the long standing
prediction that $^{146}$Ba is indeed one of the isotopes with strong
octupole collectivity \cite{footnote}.

\begin{table}
 \caption{\label{tab:me} The experimental
$|\langle I_f^\pi \| \hat{M}_{\lambda} \| I_i^\pi \rangle |$
matrix elements
($e \cdot b^{\lambda /2}$) based on the \textsc{gosia} fit along with
new symmetry-conserving configuration-mixing calculations (see text and
Ref.~\cite{Bernard2016} for details).}
 \begin{ruledtabular}
  \begin{tabular}{cccc}
$I_i^\pi \rightarrow I_f^\pi$ & $E\lambda$ & Exp. & SCCM \\
\hline
$0^+\rightarrow 1^-$ & $E1$ & 0.000223($^{10}_{-8}$) \footnote{primarily
determined by previous lifetime and/or branching ratio data
\cite{Peker1997}} & 0.00474  \\
$1^-\rightarrow 3^-$ & $E2$ & 1.2(5) & 1.6   \\
$0^+\rightarrow 2^+$ & $E2$ & 1.17(2)$^\mathrm{\; a}$ & 1.14 \\
$2^+\rightarrow 4^+$ & $E2$ & 1.97(14) & 1.90 \\
$4^+\rightarrow 6^+$ & $E2$ & 2.35($^{+20}_{-24}$) & 2.43 \\
$6^+\rightarrow 8^+$ & $E2$ & 2.17($^{+65}_{-33}$) & 2.90 \\
$0^+\rightarrow 3^-$ & $E3$ & 0.65($^{+14}_{-20}$) & 0.54 \\
$2^+\rightarrow 5^-$ & $E3$ & 1.01($^{+61}_{-20}$) & 0.87 \\
$4^+\rightarrow 7^-$ & $E3$ & 1.25($^{+85}_{-34}$) & 1.11 \\
$6^+\rightarrow 9^-$ & $E3$ & 1.5($^{+8}_{-12}$) & \\
  \end{tabular}
 \end{ruledtabular}
\end{table}

The persistence of this strong collectivity between
$^{144}$Ba$_{88}$ and $^{146}$Ba$_{90}$ confirms that the drastic
reduction in electric dipole moment between the two isotopes is not
the result of quenched octupole strength, as suggested by the
high-spin behavior of $^{146}$Ba. The sudden band alignments in
$^{146}$Ba pointed out in Ref.~\cite{Urban1997}, are then most
likely the result of a crossing between yrast and yrare bands
predicted in this mass region sometime ago
\cite{Bengtsson1979,Bengtsson1978}.
It should be noted that alternative explanations have also been proposed.
These include a transition to a more reflection-symmetric shape at moderate
spin~\cite{Nazarewicz1992}, and a description in terms of a
condensate of rotationally-aligned octupole
phonons~\cite{Frauendorf2013,Frauendorf2008}. Concerning the latter,
however, it should be mentioned that while the interleaved
sequences of opposite parity are consistent with the proposed
picture, the absence of strong $E1$ linking transitions associated
with multi-octupole phonon excitations at higher spins is not.

Over the past three decades, extensive theoretical efforts have been
devoted to understanding the variation of the $E1$ transition
strengths observed in nuclei near $^{146}$Ba and
$^{224}$Ra~\cite{Leander1986,Hamamoto1989,Peker1979,Egido1989,Butler1991,
Egido1990,Egido1992,Robledo2010,Dorso1986}.
It is generally believed that the observations are 
the result of the relation between octupole
collectivity and the non-uniform distribution of protons and
neutrons.  
This was first shown within the framework of a macroscopic-microscopic 
approach where the experimental $E1$ transition strengths could be 
described \cite{Leander1986,Butler1991}.
Early self-consistent Hartree-Fock-Bogoliubov (HFB) calculations with the 
Gogny interaction were also able to reproduce the very low values of the 
dipole moment $D_0$ in $^{224}$Ra and $^{146}$Ba~\cite{Egido1989,Egido1990}. 
However, all of these
models predicted that the nuclei under study are
reflection-asymmetric, and argued that this is at the core of the
observed variations. Thus, the recently measured strong 
octupole collectivity in $^{144}$Ba~\cite{Bucher2016} and $^{146}$Ba 
(Table~\ref{tab:me}) provides an important validation of this interpretation.

More recently, microscopic self-consistent methods have been 
improved by including beyond-mean-field correlations.
These developments provide an explanation 
of the microscopic origin of octupole
collectivity and study the impact of octupole correlations on both 
ground-state properties and electromagnetic transitions. 
To explore in $^{146}$Ba this phenomenon of a strong octupole 
collectivity accompanied
by a much reduced electric dipole moment, a
theoretical model based on mean field HFB intrinsic wave functions
has been used with a symmetry-conserving configuration-mixing method (SCCM).  
The model assumes that only the axially symmetric
quadrupole ($Q_{20}$) and octupole ($Q_{30}$) degrees of freedom 
are relevant (collectively referred to as $\mathbf{Q}$). 
A set of constrained HFB states
$|\mathbf{Q}\rangle$, subsequently projected onto good angular
momentum, parity and particle number (the corresponding states are
denoted as $|\Phi^{J,\pi} (\mathbf{Q}) \rangle $) is used as a 
variational subspace. Linear combinations of the above states
$|\Psi^{J,\pi}_\sigma\rangle = \sum_\mathbf{Q} f^{J,\pi}_\sigma
(\mathbf{Q}) |\Phi^{J,\pi} (\mathbf{Q}) \rangle$ are used in the
spirit of the generator coordinate method (GCM) to obtain the 
low-lying collective spectrum (see Ref.~\cite{Bernard2016} for a recent
account and an application to $^{144}$Ba). The interaction 
generating the intrinsic states is the well-known Gogny D1S force
\cite{berger1984}. The physics of the low-lying quadrupole and
octupole states is contained in the collective amplitudes
$F^{J,\pi}_\sigma (\mathbf{Q}) $ defined in Eq~(5) of
Ref.~\cite{Bernard2016} in terms of the GCM amplitudes $f^{J,\pi}_\sigma
(\mathbf{Q})$ (the latter are obtained by solving the Hill-Wheeler (HW)
equations of the GCM).

\begin{figure}
    \includegraphics[width=20pc]{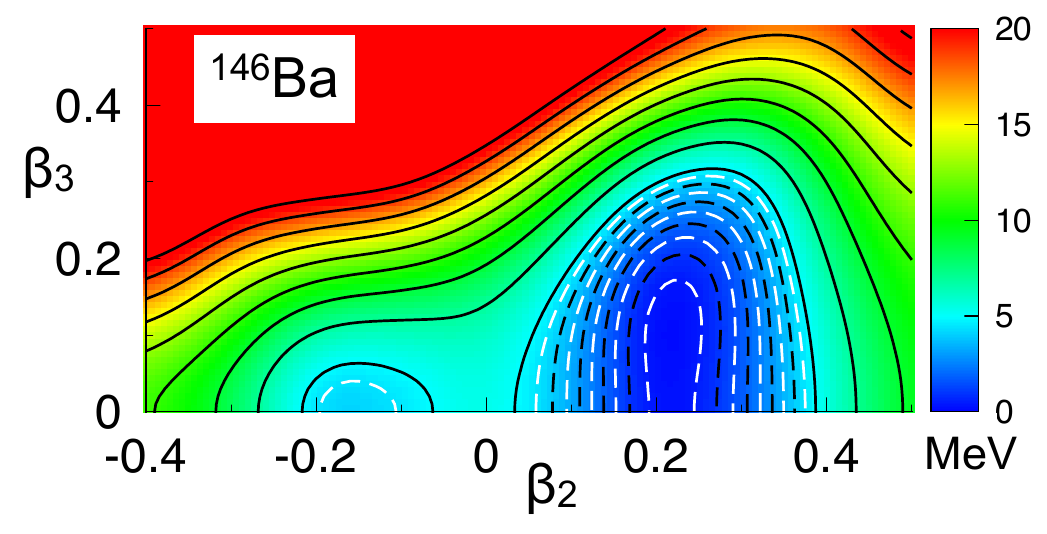}
    \caption{(Color online)
The HFB potential energy surface. Axial quadrupole ($\beta_2$) and 
octupole ($\beta_3$) deformation parameters are defined as 
$\beta_\lambda \equiv{4\pi}\langle \mathbf{q} \, | r^\lambda 
Y_{\lambda0}|\mathbf{q}\rangle / (3r^{\lambda}_{0}A^{\lambda/3+1})$
with $r_0=1.2$ fm and $A$ being the mass number.  Dashed (solid) contour 
lines are separated by 0.5~MeV (2.0~MeV).}
    \label{fig:PES}
\end{figure}

\begin{figure}
\includegraphics[width=20pc]{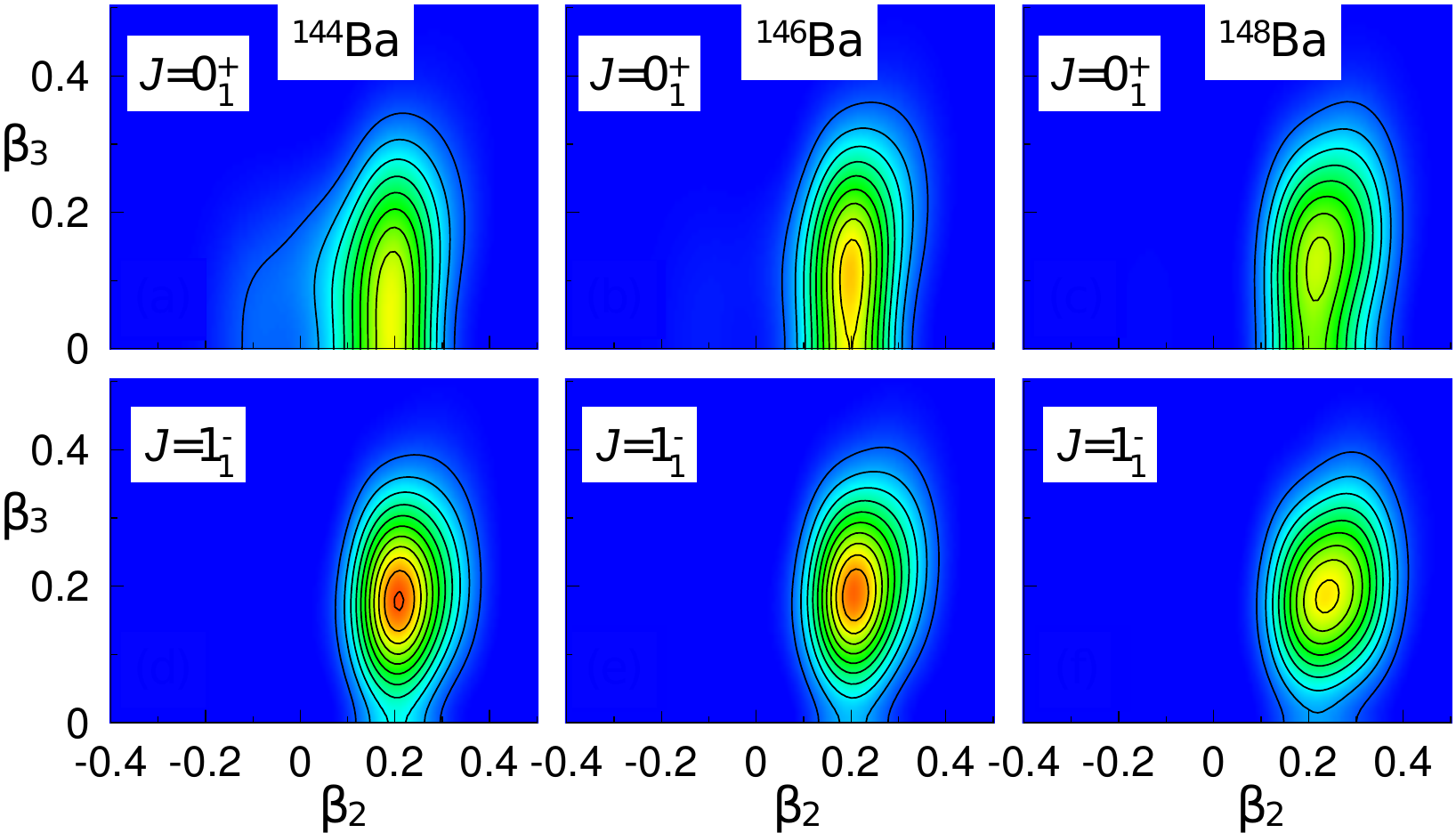}
\caption{(color online) Collective amplitudes 
corresponding to 
$^{144}$Ba (left), $^{146}$Ba (middle), and $^{148}$Ba (right).}
\label{CollWF}
\end{figure}

The HFB potential energy surface (PES) as a function of the
deformation parameters $\beta_2$ and $\beta_3$ is given in
Fig~\ref{fig:PES} for $^{146}$Ba. Note that the potential energy
is symmetric under the change of sign of $\beta_3$ due to the parity
symmetry of the nuclear interaction. A reflection-asymmetric, absolute 
minimum is obtained at $\beta_2=0.21$ and $\beta_3=\pm0.1$. The shape of
the $F^{J,\pi}$ collective amplitudes is mainly driven by the
intrinsic potential energy surface. As a consequence, the
$F^{J,\pi}$ amplitudes for $^{146}$Ba are concentrated around
those minima as seen in Fig~\ref{CollWF} where they are compared with 
those for $^{144}$Ba and $^{148}$Ba. In the three nuclei, the $1^-$ and
$0^+$ collective amplitudes have a very large overlap,
characteristic of strong octupole correlations.  The energy of the $1^-$ 
state is well-reproduced, however the ground-state and negative-parity 
sequences are characterized by a smaller moment of inertia than observed, 
due to limitations discussed in Ref.~\cite{Bernard2016}.

The electromagnetic transition strengths have been computed without
invoking effective charges or uncontrolled approximations. The
calculated values are compared with the experimental data in
Table~\ref{tab:me}. There is fair agreement between the calculated
and measured $E\lambda$ matrix elements, including the strong
$B(E3)$ strengths between the $3^-$ and $0^+$ and the much quenched
$E1$ between the $1^-$ and $0^+$ states. In the present microscopic
framework, the $B(E1)$ transition strength is proportional to the
square of the overlap of the dipole operator between the initial
$1^-$ and final $0^+$ states. There are two basic ingredients
entering the required overlap (see Eq~(4) of
Ref.~\cite{Bernard2016}). One is the structure of the collective
wave functions $F^{J,\pi}_\sigma(\mathbf{Q})$, and the other is the
overlap between the projected intrinsic states $|\Phi^{J,\pi}
(\mathbf{Q}) \rangle$. Because the values of
$F^{J,\pi}_\sigma(\mathbf{Q})$, as shown in Fig~\ref{CollWF},
exhibit little variation with neutron number in the three Ba
isotopes, the sudden drop of $B(E1)$ in $^{146}$Ba has to be
associated with the overlap of the dipole operator between the
projected intrinsic states $|\Phi^{J,\pi} (\mathbf{Q}) \rangle$.
Specifically, the calculations indicate that the dipole moment
changes from positive values in $^{144}$Ba, to nearly zero values in
$^{146}$Ba, and finally to negative values in $^{148}$Ba.  Hence, the 
changes in $E1$ strengths with neutron number are associated with changes 
in the intrinsic dipole moment linked to the evolving mean field.  A similar 
conclusion was reached in Ref.~\cite{Egido1990}.

\begin{figure}
\includegraphics[width=20pc]{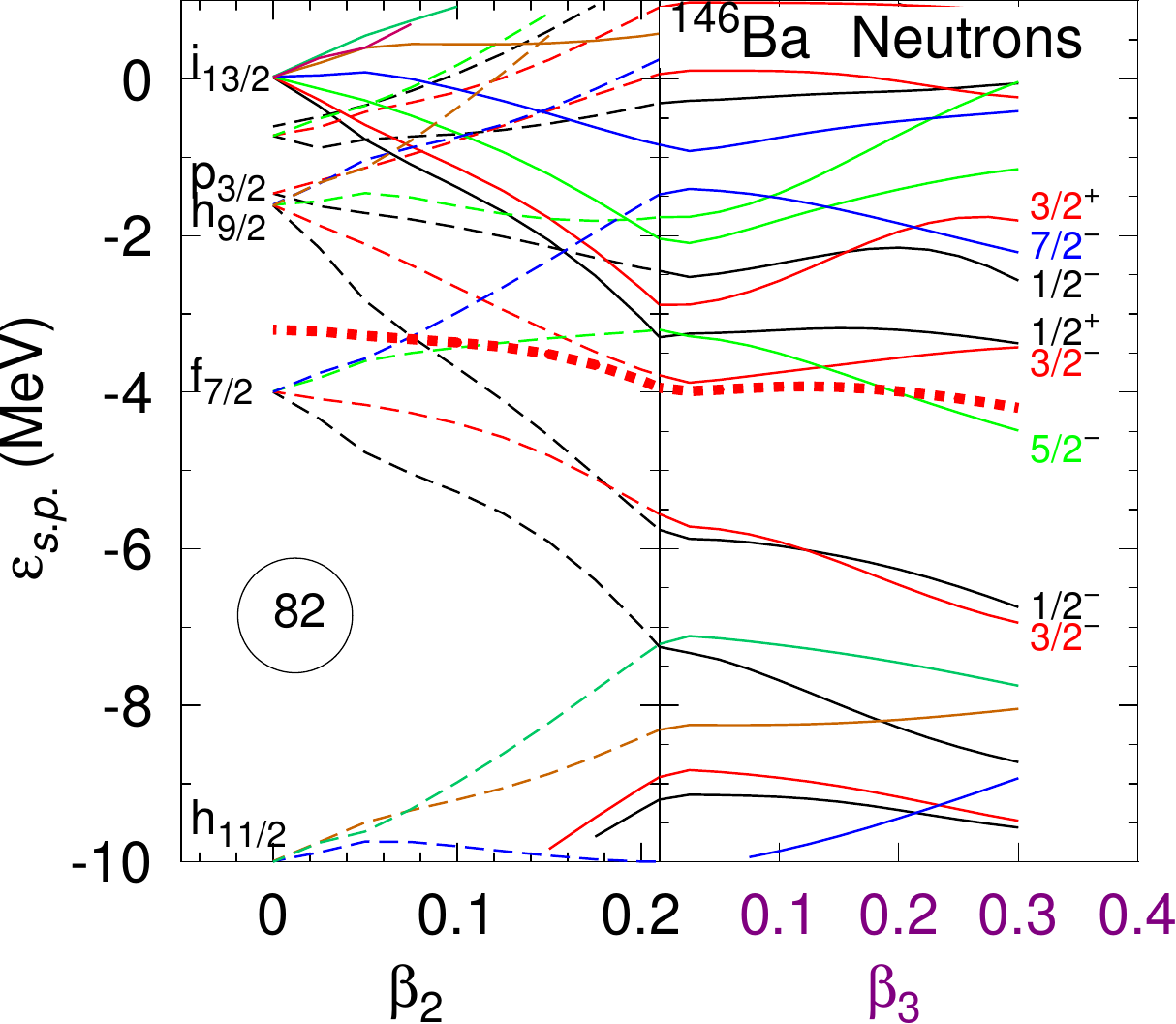}
\caption{(Color) Neutron single-particle energies (SPE) as a 
function of
$\beta_2$ (left) and as a function of $\beta_3$ (right) for $^{146}$Ba. 
The right panel is calculated with a constant value of $\beta_2$=0.2 
(ground-state value) starting with $\beta_3$=0 at the central vertical axis. 
A thick dotted line shows the Fermi level. 
The SPE as a function of $\beta_2$ is used
to justify the spherical orbital assignments and parities of the 
relevant neutron
orbitals (see text for details). }
\label{Spe}
\end{figure}

The behavior of the dipole moment with neutron number in
these Ba isotopes can further be traced back to the occupation of specific
single-particle states near the Fermi surface. 
Considering the
evolution of the single-particle energies with $\beta_3$ 
in Fig~\ref{Spe}, three neutron orbitals are of interest with $K^{\pi}$ 
quantum numbers $3/2^-$, $5/2^-$ and
$1/2^+$. These are of $h_{9/2}$, $f_{7/2}$, and
$i_{13/2}$ spherical parentage, respectively. The three states are empty in
$^{144}$Ba, but have significant respective occupancies of 
$v^2=0.4$, $v^2=0.27$, and $v^2=0.20$
in $^{146}$Ba.  As visualized in Fig.~4 of Ref.~\cite{Egido1990}, their 
occupation results in a contribution to the dipole moment which almost 
cancels that by the protons.  As the occupancies increase further with two 
additional neutrons in $^{148}$Ba, the (total) dipole moment changes sign 
and returns to a sizeable value.

To summarize, the $E3$ strength in short-lived $^{146}$Ba was measured 
directly by 
multi-step Coulomb excitation with GRETINA and CHICO2.  The longstanding 
prediction of an 
enhanced octupole collectivity was verified. 
The data also provide firm experimental evidence that the large drop of 
$B(E1)$ value is not the result of quenched octupole collectivity in 
$^{146}$Ba.  
Such a collectivity is well-reproduced by the SCCM model with the 
Gogny energy density functional, and the variation in $E1$ strength between 
isotopes is associated with changes in the neutron occupancy  
of high-$j$, low-$K$ orbitals located near the Fermi 
surface.  The present results help validate the general character of the 
microscopic origin of large variations in electric dipole moments in 
the reflection-asymmetric nuclear potential, and they represent an important 
confirmation of such effects in the Ba region of neutron-rich nuclei.

This material is based upon work supported by the U.S. Department of Energy, 
Office of Science, Office of Nuclear Physics under Contract No. 
DE-AC02-06CH11357 (ANL).  Work at LLNL and INL is supported by the U.S. 
DOE under respective Contracts No. DE-AC52-07NA27344 and No. 
DE-AC07-05ID14517.  GRETINA was funded by the U.S. DOE-Office of 
Science, Office of Nuclear Physics by the ANL contract No. above and by 
Contract No. DE-AC02-05CH11231 (LBNL). 
This research used resources of ANL's ATLAS facility, which is a DOE Office
of Science User Facility. The  work of LMR was supported by the
Spanish grants FIS2012-34479-P MINECO, FPA2015-65929-P MINECO and 
FIS2015-63770-P MINECO, and TRR by the Spanish grants  
FIS-2014-53434-P MINECO and Programa Ram\'on y Cajal 2012 No. 11420.

\bibliography{Ba146LettRefs}

\end{document}